\documentclass[pteplogo,hdvipdfmx]{ptephy_v1}

\preprintnumber{XXXX-XXXX} 

\usepackage{amsmath,amssymb}
\usepackage{graphicx}
\usepackage{subfigure}
\usepackage{bm}
\usepackage{ulem}
\usepackage{slashed}
\usepackage{dcolumn}
\usepackage{epsfig}
\usepackage{multirow}
\usepackage[all, warning]{onlyamsmath}

\newcommand{\del}[2]{\frac{d #1}{d#2}}

\renewcommand{\vec}[1]{{\bm #1}}

\usepackage{color}

\allowdisplaybreaks

\begin{document}


\title{Excitation spectra of heavy baryons in diquark models}


\author[1,2]{Kento Kumakawa}%
\author[2,1]{Daisuke Jido}%
\affil{Department of Physics, Tokyo Metropolitan University,  Hachioji, Tokyo, 192-0397, Japan}
\affil{Department of Physics, Tokyo Institute of Technology, Meguro, Tokyo 152-8551, Japan}




\begin{abstract}
The excitation energy spectra of heavy baryons consisting of a heavy quark and two light quarks
are investigated by using diquark models in order to examine the nature of the diquark 
as a constituent of single heavy baryons. 
We consider two diquark models; in model A the diquark is treated as a point-like particle, 
while it has a spatial size in model B. We determine the masses of scalar and axial vector diquarks by
the mass difference of the ground state charmed baryons to the $\Lambda_{c}$ baryon, while
the mass of the $ud$ scalar diquark in the $\Lambda_{c}$ baryon is assumed to be 500 MeV as a
reference. The parameters of these models are fixed by the $1p$ excitation energy of $\Lambda_{c}$.   
We find that model A reproduces well the excitation energy spectra of the charmed and bottomed baryons, 
although the string tension of the confinement potential in model A should be a half of that of the charmonium, 
while Model B suggests degeneracy of the $2s$ and $1d$ states, which is not seen in the $\Lambda_{c}$ spectrum. 
\end{abstract}

\maketitle

\section{Introduction}

Understanding the hadron structure in terms of dynamics of quarks and gluons is one of the
most important issues in hadron physics. For this purpose, the quest of effective constituents 
of hadrons is a clue to understand the structure of the hadrons. The constituent quarks are 
such effective degrees of freedom, explaining, for instance, the fine structure of heavy 
quarkonia~\cite{Eichten:1974af,Mukherjee:1993hb} and the static properties, 
such as the magnetic moments
of the light baryons~\cite{Isgur79,Capstick94}.
Along this line we try to find the possibility that the diquark also can be a constituent of hadrons. 

The diquark is a pair of quarks and was already mentioned in Ref.~\cite{Gell-Mann:1964ewy}
at the same time when quarks were introduced to explain the spectra of 
hadrons~\cite{Gell-Mann:1964ewy,Zweig:1964ruk}.
The diquark is a colored object and cannot be observed directly in experiments.
Since introduced as a constituent of baryons~\cite{Ida:1966ev,Lichtenberg:1967zz},  
the diquark correlations have been pointed out so far phenomenologically
as summarized in Ref.~\cite{Anselmino:1992vg,Jaffe:2004ph}. Among various diquarks,
one expects stronger attraction between two quarks in flavor, spin and color antisymmetric
configurations, which is so-called good diquark, owing to the most attractive color magnetic 
interaction in perturbative calculation and instanton-induced interaction~\cite{Jaffe:2004ph}. 
The good diquark is also favored
in lattice calculation~\cite{Hess98,Babich,Orginos,Alexandrou}. 
The characteristic size of the diquark was suggested to be about $1.1 \pm 0.2$ fm 
in lattice calculation~\cite{Alexandrou} and to be smaller than 1 fm in the Schwinger-Dyson
approach~\cite{Imai:2014yxa}. 
It has been reported that light scalar mesons are described well in 
a diquark-antidiquark picture~\cite{Jaffe:1976ig,Black:1999yz,Maiani:2004uc,tHooft:2008rus}
and their strong decays are reasonably reproduced in 
Refs.~\cite{Maiani:2004uc,tHooft:2008rus}.
The diquark has been also studied extensively in the quark model approaches~\cite{Goldstein:1979wba,Lichtenberg:1969sxc,Lichtenberg:1982jp,Liu:1983us,Ebert:2005xj,Ebert:2007nw,Hernandez:2008ej,Lee:2009rt,Wang:2017vnc}, and 
recently it has been found in Refs.~\cite{Jido:2016yuv,Kumakawa:2017ffl} 
that the confinement potential for the heavy quark and the light diquark 
should be weaker than that for the heavy quark and antiquark in quarkonia 
in order to produce the first excitation energy of the $\Lambda_{c}$ baryon. 
A diquark QCD sum rule was proposed in Ref.~\cite{Kim:2011ut} and  
found the constituent $ud$ diquark mass to be around 0.4~GeV for 
the $\Lambda$, $\Lambda_{c}$ and $\Lambda_{b}$ baryons. 
Recently a symmetry between the $\bar s$ quark and the $ud$ diquark was
introduced in Ref.~\cite{Amano:2019jek} based on the V(3) symmetry~\cite{miya66,miya68}
and proposed a new classification scheme of the heavy hadrons. 
Based on this symmetry,
a sum for the weak decay rates of $\bar B_{s}^{0}$ and $\Lambda_{b}$ was derived in Ref.~\cite{Amano:2021spn}.
A quark-diquark symmetry for heavy systems was also discussed in Ref.~\cite{Nielsen:2018uyn}.

Thanks to the strong correlation between light quarks, the diquark may play an important 
role of the structure of baryons. If so, diquarks can be effective constituents of hadron.
Nevertheless, as discussed in Ref.~\cite{Jaffe76-1}, 
tetraquark meson wavefunctions are found to be dominated by quark-antiquark configurations, which tells us that diquark correlations may be rather suppressed even in multiquark hadrons. This also implies that the diquark correlation is possibly fragile and 
{the reformation of the quark-antiquark pairs from the diquark and anti-diquark
can be important in light quark systems.}
Therefore, baryons with a single heavy quark are good systems to 
investigate the effectiveness of diquark as a constituent of hadrons.
If the diquark picture works for the single heavy baryons,
their excitation spectra should be explained by two-body dynamics of a heavy quark and a diquark.

In this paper, 
we investigate the properties of the diquark constituents in the hadron structure 
and examine whether the constituent diquark picture works well. For this purpose, 
we perform a phenomenological investigation of the excitation energy spectra of the single 
heavy baryons, $\Lambda_{c}$, $\Xi_{c}$, $\Sigma_{c}$, $\Omega_{c}$ 
and the counterparts of the bottomed baryons. 
Here we assume that these baryons are composed of a light diquark and a heavy quark.
It is not our intention that we would reproduce the precise energy spectra of 
these baryons grounded on fundamental theories, but we would rather grasp the global feature 
of the heavy baryons from the observed spectra using simple models. 
For this reason, we focus on the orbital excitations 
in the confinement potential and pay no attention to the fine structure induced by 
the spin dependent forces.
Here we take two diquark models; Model A considers a point-like diquark
with a weaker string tension than the heavy quarkonium systems as suggested in 
Ref.~\cite{Jido:2016yuv},
while Model B introduces a spatial size of the diquark for the color interaction between 
the heavy quark and the diquark by following Ref.~\cite{Kumakawa:2017ffl}.

The paper is organized as follows. In Sec.~\ref{sec:formulation}, we explain our models, and
construct the two-body potential of the heavy quark and diquark system. 
In Sec.~\ref{sec:para} we determine the model parameters by the first excitation energy 
of the $\Lambda_{c}$ baryon 
and the diquark masses other than the $ud$ scalar diquark by the ground state masses
of the charmed baryons.
In Sec.~\ref{sec:results}, we show our numerical results. 
After showing the excitation spectrum of $\Lambda_{c}$ in Sec.~\ref{sec:Lambdac}, we
show the calculated excitation spectra of the heavy baryons
in Sec.~\ref{sec:spectra}. We also discuss the Regge trajectory of the obtained mass spectra 
in Sec.~\ref{sec:Regge} and the excitation energy of the doubly charmed baryon $\Xi_{cc}$ in 
Sec.~\ref{sec:double}. Finally, Sec.~\ref{sec:summary} is devoted to summary and conclusion.

\section{Formulation}
\label{sec:formulation}

In this paper, we treat heavy baryons as two-body systems of a light diquark and a 
heavy quark. The light diquarks that we consider here are composed of two quarks 
out of the $u$, $d$ and $s$ quarks and form color anti-triplet. 
The spin of the diquark is either singlet ($S=0$) for the flavor antisymmetric 
configuration or triplet ($S=1$) for the flavor symmetric case. 
We treat the diquark as a fundamental degree of freedom and do not consider 
radial excitations of two quarks in the diquark nor
spin transitions of the diquark induced by rotational excitations of the diquark.
For the heavy quark $h$ we consider the charm and bottom quarks. 

In Table~\ref{tab:diquark}, we summarizes the heavy baryons that we consider in this work
and also show the quantum numbers of the diquark appearing in the heavy baryon.
The $\Lambda_{h}$ baryon has isospin $I=0$ and is composed a $[ud]$ isoscalar-scalar 
diquark and a heavy quark. 
The $\Xi_{h}$ baryon has isospin $I=1/2$ with one strange quark, and 
is made up with a $qs$ diquark and a heavy quark. 
Here we call $u$ and $d$ quarks by $q$ collectively. 
The spin of the $qs$ diquark can be singlet or triplet, and thus
we distinguish the $\Xi_{h}$ baryon by the diquark spin:
$\Xi_{h}$ has the spin 0 diquark $[qs]$ of which flavor configuration is antisymmetric 
under the quark exchange, while $\Xi_{c}^{\prime}$ has the spin 1 $\{qs\}$ diquark with 
symmetric flavor configuration. 
These $\Xi_{h}$ baryons have the same quark contents.
To distinguish them experimentally, one has to understand their diquark structure. 
The $\Sigma_{h}$ baryon has isospin $I=1$ and is constructed by a spin 1 $\{qq\}$
diquark and a heavy quark. Finally $\Omega_{h}$ has two strange quarks and 
is composed of a spin 1 strange diquark, $\{ss\}$, and a heavy quark.

 \begin{table}[htb]
  \caption{Diquark quantum numbers in the heavy baryons. 
  The light quark $q$ denotes the up or down quark.
  The bracket $[\cdots]$ and brace $\{\cdots\}$ in the flavor row
  stand for symmetric and antisymmetric under the exchange of two quarks, respectively. 
  }
    \label{tab:diquark}
    \centering
    \begin{tabular}{r|ccccc} 
    \hline\hline
       heavy baryon & $\Lambda_h$ & $\Xi_h$ & $\Sigma_h$ & $\Xi^{'}_h$ & $\Omega_h$  \\ 
       \hline
        isospin $(I)$ & $0$ & $1/2$ & $1$ & $1/2$ & $0$  \\ 
        spin $(S)$ & 0 & 0 & 1 & 1 &1 \\
         flavor $(f)$& $[ud]$ & $[qs]$ & $\{qq\}$ & $\{qs\}$ & $\{ss\}$ 
    \\ \hline\hline
    \end{tabular}
\end{table}

\subsection{Schr\"odinger equation}
We compare two diquark models. In Model A, the diquark is assumed to be a point-like particle. This model was investigated in Ref.~\cite{Jido:2016yuv} for the $\Lambda_{c}$ baryon. In the second model, Model~B, we consider the size of the diquark to evaluate the interaction between the heavy quark and diquark. 
Reference~\cite{Kumakawa:2017ffl} introduced the size of the diquark by treating the diquark as a rigid rotor made up of two quarks spatially separated with a fix distance. Here we use a Gaussian distribution for the size of the diquark, which is more suitable for quantum systems.  
In both models, we consider the heavy baryons as a two-body system of a heavy quark and a diquark. 
The Hamiltonian in the center of mass system 
is given as 
\begin{equation}
   \hat H = m_{h} + m_{d}
    - \frac{1}{2\mu}\frac{1}{r}\del{^{2}}{r^{2}} 
   + \frac{\hat {\vec L}_{r}^{2}}{2\mu r^{2}}  + V( r),  \label{eq:H}
\end{equation}
where $r$ is the distance between the diquark and heavy quark, 
$m_{h}$ and $m_{d}$ are the masses of 
the heavy quark  and the diquark, respectively, 
$\mu$ is the reduced mass given by $\mu = m_{h} m_{d}/(m_{h} + m_{d})$,
$\hat{\vec L}_{r}$ is the orbital angular momentum operator 
of the two-body system, and $V(r)$ is the two-body potential for the heavy quark and the diquark.
The potential is commonly used in the whole calculations for the heavy baryon spectra
once the model parameters are fixed.
The explicit form of the potential depends on the model. 
We take a central potential and do not consider fine structure splittings 
induced by the spin-spin, spin-orbit ant tensor interactions 
in order to investigate global structure of excitation spectra. 

The total angular momentum of the system reads $\vec J = \vec S + \vec L$ with
the total spin $ \vec S = \vec S_{h} + \vec S_{d}$ given by
the heavy quark spin $\vec S_{h}$ and the diquark spin $\vec S_{d}$
and the total orbital angular moment $ \vec L = \vec L_{r} + \vec L_{\rho}$ where 
$\vec L_{\rho}$ is the diquark rotational angular moment for Model~B. 
The spin of the heavy quark is 1/2, while that of the diquark is 0 or 1 
depending on the flavor of diquark. 
Because we do not consider the fine structure of the energy spectra in the present work, 
the baryon masses can be labeled by the orbital angular momentum $L$. 
Thus, the angular part of the Schr\"odinger equation with the Hamiltonian~\eqref{eq:H} is 
solved with the spherical harmonics~$Y_{L}^{M}$,
where $M$ is the magnetic quantum number for the orbital angular momentum $\vec L$.

To obtain the energy of each state with the angular momentum~$L$, $E_{L}$,
we just solve the radial differential equation:
\begin{equation}
   \left[- \frac{1}{2\mu} \frac{d^{2}}{dr^{2}} + \frac{ L(L+1)}{2\mu r^{2}} + V(r) \right] \chi_{L}(r) = 
   E_{L} \chi_{L}(r),  \label{Scheq}
\end{equation}
where the radial wavefunction $\chi_{L}(r)$ is normalized as 
$$
\int |\chi_{L}(r)|^{2} dr = 1.
$$ 
We assume the flavor symmetry among the diquarks and use a common interaction 
potential~$V(r)$ for the heavy baryons. The symmetry breaking appears in the diquark mass.

\subsection{Model A}

In Model A the diquark is treated as a point-like particle.  
{The main component of the confinement potential may be brought about by 
the color electric interaction, which primarily depends on the color charge and is insensitive to the masses.}
Thanks to the same color structure of the quark-diquark systems as that of the  quark-antiquark systems, 
we take the interaction potential from the quarkonium systems. 
Following Ref.~\cite{Jido:2016yuv},
here we use the so-called Coulomb-plus-linear potential \cite{Eichten:1974af} given by
\begin{equation}
   V_\textrm{A}(r) = - \frac 43 \frac{\alpha_{s}}{r} + k r + V_{0}  \label{eq:potA}
\end{equation}
with the strength of the Coulomb potential, $\alpha_{s}$, and the string tension $k$ are the model parameters. The parameter $V_{0}$ determines the absolute value of the baryon mass.
According to the universality of the color interaction, the color electric potential $V(r)$ should be common in the quark-antiquark and quark-diquark systems. The values which reproduce the global structure of the excitation spectra of charmonium and bottonium were found to be $\alpha_{s} \simeq 0.4$ and $k \simeq 0.9$ GeV/fm in Ref.~\cite{Jido:2016yuv,Quigg:1979vr}. This parameterization reproduces well the excitation spectra of the open charm and bottom systems~\cite{Jido:2016yuv}. Contrary to our expectation, it was found in Ref.~\cite{Jido:2016yuv} that these values do not reproduce the $\Lambda_{c}$ excitation spectrum and we have to use a half strength of the string tension. Thus, we determine the strength of the string tension later by the excitation energy of $\Lambda_{c}$.

\subsection{Model B}
\label{sec:model_B}

Also in Model B, we treat the diquark as a kinematically point-like particle, 
but we introduce the size and deformation of the diquark.
We consider the diquark as a rigid rotor with a size $\rho$
and calculate the interaction between the heavy quark and the rigid rotor.
Folding this interaction with a Gaussian distribution for the diquark size
we obtain an effective potential between the heavy quark and the diquark. 
With the size of the diquark, we consider its rotational excitations.
According to symmetry of quarks in the diquark,
the rotational angular momentum of the diquark must be an even number.
This is because the diquark has color anti-triplet, which is antisymmetric 
under the quark exchange, and we consider symmetric spin-flavor configurations 
for the diquarks. Thus, the spacial configuration of the quarks in the diquark must be 
symmetric under the quark exchange. Because the parity of the states with angular momentum 
$\ell$ under spacial inversion is given as $(-)^{\ell}$, the rotational angular momentum
of the diquark considered here must be an even number.
We consider the rotational excitation with $L_{\rho}=2$
for low-lying states. The diquark with $L_{\rho}=2$ has deformation. 
One might consider that $L_{\rho}=1$ excitation with the spin rearrangement 
of the diquark would have a comparable excitation energy with that of 
the $L_{\rho}=2$ without the spin rearrangement. These kinds of excitation 
can be described fully in quark models in which light quarks are treated individual 
particles. Since we consider a diquark model in order to examine the nature 
of the diquark as an effective constituent of hadrons here,  
we do not consider the spin rearrangement of the diquark. 

The radial coordinate $r$ in the Sch\"odineger equation \eqref{Scheq} 
denotes the distance between the heavy quark and the center of the diquark. 
The total orbital angular momentum of the system is given by 
$\vec L = \vec L_{r} + \vec L_{\rho}$.
In general, a state with total orbital angular momentum~$L$ is composed of several 
combinations of $L_{r}$ and $L_{\rho}$.
According to Ref.~\cite{Kumakawa:2017ffl}, the lowest lying  
$L=0$ and $L=1$ states are found to be constructed essentially by the states 
with $(L_{\rho}, L_{r}) = (0,0)$ and $(0,1)$, respectively.
This is because the rotational excitation needs two quanta and 
costs more than the orbital excitation $L_{r}$. 
Thus, we take the $(L_{\rho}, L_{r}) = (0,0)$ and $(0,1)$ configurations 
for the lowest $S$ and $P$ states, respectively.
For the $D$ state $(L=2)$, the states with 
$(L_{\rho}, L_{r}) = (0,2)$ and $(2,0)$ may have a comparable energy depending on the size of 
the diquark. As we will find later, with the diquark size appropriate to reproduce the $1p$ excitation 
energy, these two states have a very similar energy and can be mixed.
Here, we consider the $(0,2)$ and $(2,0)$ configurations for $L=2$
and diagonalize their admixture. For these states we call the one 
which has a larger contribution of the $(0,2)$ configuration 
the $d_{\lambda}$ state and the other the $d_{\rho}$. 

We follow Ref.~\cite{Kumakawa:2017ffl} to calculate the effective potential 
$v(r; \rho)$ for the interaction between a heavy quark and a fixed-size diquark.
The interaction between the quarks forming the color anti-triplet configuration, $v_{qq}(\vec r)$,
is assumed to be half of the color singlet quark-antiquark interaction $v_{\bar qq}(\vec r)$ and
the $\bar qq$ potential is described by the Coulomb-plus-linear type form:
\begin{equation}
   v_{qq}(\vec r) = \tfrac12 v_{\bar qq}(\vec r) = \frac12 \left[-\frac43 \frac{\alpha_{s}}{r} + k^{\prime} r \right].
   \label{eq:intB}
\end{equation}
The parameters, $\alpha_{s}$ and $k^{\prime}$, are determined so as to reproduce the charmonium 
and bottonium spectra~\cite{Jido:2016yuv} as
\begin{equation}
   \alpha_{s}  = 0.4, \qquad k^{\prime} = 0.9\ \textrm{[GeV/fm]}. 
\end{equation}
The light quarks in the diquark are apart off distance~$\rho$. 
The interaction between the heavy quark and the light quarks is written as
\begin{equation}
   v_{hd}(\vec r, \vec \rho) = v_{qq}(\vec r - \tfrac12 \vec\rho) + v_{qq}(\vec r + \tfrac12 \vec \rho),
\end{equation}
where we take the geometrical middle point between two light quarks as the center of the diquark. 
Projecting out the component of the total orbital angular momentum $L$ composed of $L_{r}$ and $L_{\rho}$
for the potential $v_{hd}(\vec r, \vec \rho)$, we obtain
\begin{align}
   v(r; \rho) = & \int d\Omega_{r} d\Omega_{\rho} 
   \left[Y_{L_{\rho}^{\prime}}^{M_{\rho}^{\prime}}(\Omega_{\rho}) 
   Y_{L_{r}^{\prime}}^{M_{r}^{\prime}}(\Omega_{r})\right]_{L}^{*}  
   v_{hd}(\vec r, \vec \rho) \left[Y_{L_{\rho}}^{M_{\rho}}(\Omega_{\rho}) 
   Y_{L_{r}}^{M_{r}}(\Omega_{r})\right]_{L},   \label{fixedrho}
\end{align}
where $[\cdots]_{L}$ is understood to take the appropriate linear combination to compose 
the total angular momentum $L$ in terms of $L_{r}$ and $L_{\rho}$. 
The details of the derivation of the potential~\eqref{fixedrho} is explained in Ref.~\cite{Kumakawa:2017ffl}.

We consider a Gaussian distribution for the size of the diquark. 
The effective potential between the heavy quark and the diquark for model B, $V_\textrm{B}(r)$, 
is obtained by folding the potential $v(r; \rho)$ calculated in Eq.~\eqref{fixedrho}
with the Gaussian form of the size distribution of the diquark:
\begin{equation}
   V_\textrm{B}(r) = \int_{0}^{\infty} v(r;\rho) \sigma(\rho;\beta) \rho^{2} d\rho + V_\textrm{rot} + V_{0}, \label{eq:potB}
\end{equation}
where 
a constant $V_\textrm{rot}$ represents the rotational energy of the diquark,
the parameter $V_{0}$ determines the absolute value of the baryon mass.
The size distribution $\sigma(\rho;\beta)$ with a Gaussian parameter $\beta$ is given by
\begin{equation}
   \sigma(\rho; \beta) = \left\{ \begin{array}{cl} 
   \displaystyle \frac{4}{\beta^{3}\sqrt \pi} \exp \left( - \frac{\rho^{2}}{\beta^{2}} \right) 
   & \ \textrm{for} \ \   L_{\rho} = 0, \\
   \displaystyle \frac{16}{15 \beta^{7}\sqrt \pi} \rho^{4}\exp \left( - \frac{\rho^{2}}{\beta^{2}} \right) 
   & \ \textrm{for} \ \   L_{\rho} = 2.
   \end{array} \right.  \label{eq:gauss}
\end{equation}
The rotational energy of the diquark is evaluated as
\begin{equation}
   V_\textrm{rot} = \int_{0}^{\infty} \frac{L_{\rho}(L_{\rho}+1)}{2I(\rho)} \sigma(\rho;\beta) \rho^{2} d\rho,
\end{equation}
with the moment of inertia $I(\rho) = \rho^{2} m_{d}/4$. For $L_{\rho}=2$, the 
rotational energy is calculated as
$V_\textrm{rot} = 24/(5 \beta^{2} m_{d}) $.

\section{Parameter determination}
\label{sec:para}

In this section, we first determine the model parameters appearing in the potential of 
each model by reproducing the $1p$ excitation energy of the $\Lambda_{c}$ baryon.
In the determination of the model parameters,
we use the masses of the $ud$ diquark and the charm quark as $0.5$ GeV/c$^{2}$ and 
$1.5$ GeV/c$^{2}$ as an example for the purpose
to discuss the global feature of the heavy baryon spectra in the diquark models.
For more detailed discussion, one can perform fine-tuning of the model parameters 
including these masses so as to reproduce all of the baryon masses. 
After fixing the model parameters by the $\Lambda_{c}$ spectrum, 
we apply the same potential for the other heavy baryons, because we consider
the interaction between the heavy quark and the diquark to be universal. 
The diquark masses are determined so as to produce the masses of the ground state 
charmed baryons with the potential determined by the $\Lambda_{c}$ spectrum.

\subsection{Model parameter determination by $\Lambda_{c}$ spectrum}
The model parameters are determined so as to reproduce the $1p$ excitation energy of
the $\Lambda_{c}$ baryon. 
To calculate the $1p$ excitation energy,
we use the spin-weighted average mass of the first excited states,
$\Lambda_{c}(2595)$ with spin $1/2^{-}$ and $\Lambda_{c}(2625)$ with spin $3/2^{-}$,
because we do not consider the fine structure induced by the spin dependent interactions.
This average removes the effect of the spin-orbit interaction perturbatively. 
The $1p$ excitation energy is given by
\begin{equation}
   E_{1p}^\textrm{obs} = \frac {m_{\Lambda_{c}(2595)} + 2 m_{\Lambda_{c}(2625)}}3 - m_{\Lambda_{c}} 
   = 0.33\ \textrm{GeV}.
   \label{eq:1pLc}
\end{equation}

\subsubsection{Model A}

It has been reported that the string tension which reproduces the spectra of
the charmonium and bottonium is not suitable for $\Lambda_{c}$ and $\Lambda_{b}$
and their excitation energies are obtained with a half strength of the 
string tension~\cite{Jido:2016yuv}. In this work 
we take the value of the string tension as $k=0.5$ GeV/fm. 
A recent lattice calculation~\cite{Watanabe:2021nwe} has suggested that the 
Coulomb attraction should be considerably smaller in the diquark-quark potential than 
that in the quark-antiquark system and this might be attributed to the size of the diquark.

\begin{figure}
\centering
\includegraphics[width=0.7\linewidth]{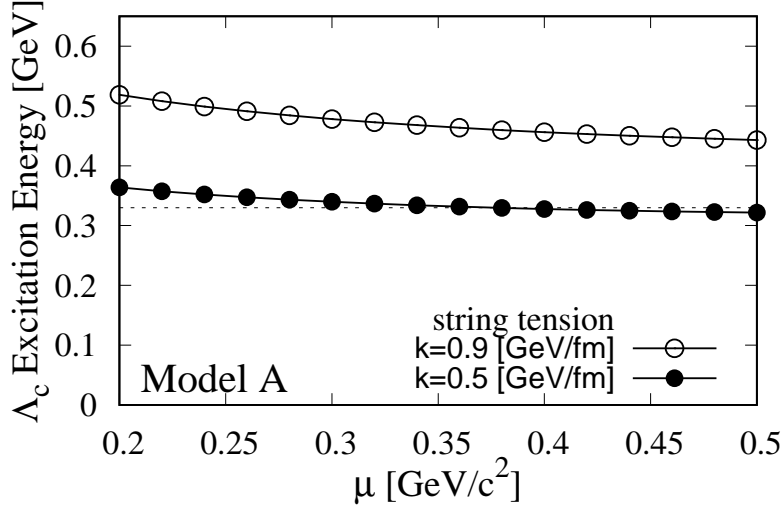}
\caption{Excitation energies of the $1p$ state of the $\Lambda_{c}$ baryon
calculated by Model A with $k=0.9$ and $k=0.5$ GeV/fm 
as functions of the reduced mass $\mu$ of the scalar $ud$ diquark and the charm quark.
The horizontal dotted line stands for the spin averaged excitation energy
given in Eq.~\eqref{eq:1pLc}. 
The energies shown in the figure are measured from the $1s$ ground state.}
\label{fig:LcAmu}
\end{figure}

One may wonder whether there is room to reproduce the $\Lambda_{c}$ excitation energy 
by adjusting the masses of the diquark and charm quark within their ambiguities. 
The value of the scalar $ud$ diquark mass may be in the range of 0.3 to 0.6~GeV
and the possible value of the charm quark mass in quark models may be within 1.27 to 1.6~GeV.
For these values the reduced mass of the diquark and charm quark is in the range 
of 0.2 to 0.5~GeV. We calculate the $1p$ excited energies 
both for $k=0.9$ and $0.5$ GeV/fm as functions of the reduced mass~$\mu$. 
Figure~\ref{fig:LcAmu} shows the values of the excited energies 
measured from the $1s$ ground state energy.  
The horizontal dotted line stands for the spin averaged excitation energy given
in Eq.~\eqref{eq:1pLc}. 
The figure shows that the calculation with the string tension $k=0.9$ GeV/fm overestimates 
largely the observed excition energy in the wide range of the reduced mass. Therefore,
the string tension which reproduces the quarkonium spectra is not suitable for the 
quark-diquark interaction in the charmed baryon. 
In addition, the decay property of $\Lambda_{c}$ also prefers $k=0.5$~GeV/fm~\cite{Jido:2016yuv}. 

Determining the absolute value of the potential energy so as to reproduce 
the mass of the ground state of the $\Lambda_{c}$ baryon, we obtain 
\begin{equation}
   V_{0} = - 0.054\ \textrm{GeV},
\end{equation} 
for Model A. 

\subsubsection{Model B}
\label{sec:ModelBpara}
\begin{figure}
\centering
\includegraphics[width=0.7\linewidth]{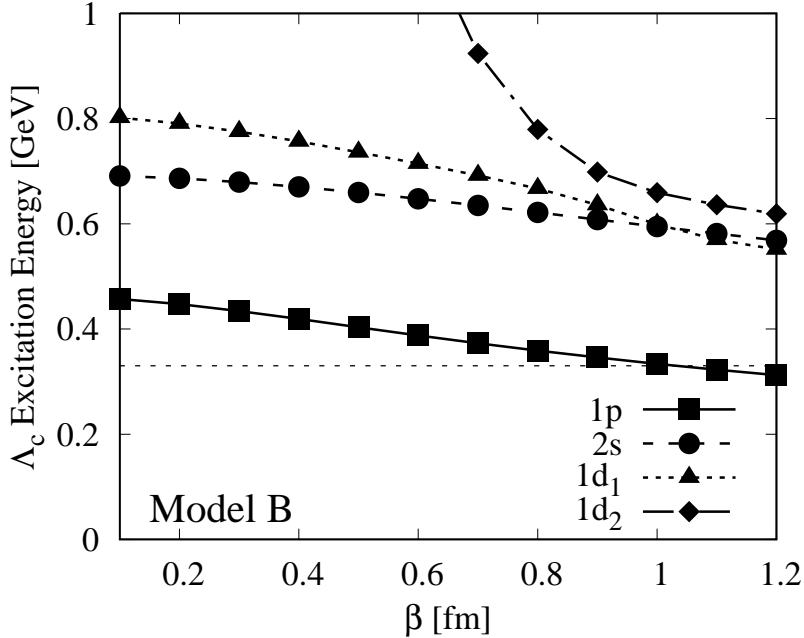}
\caption{Excitation energies of the $\Lambda_{c}$ baryon in Model B as functions 
of the Gaussian parameter $\beta$. 
The energies are measured from the corresponding $1s$ ground state. 
The horizontal dotted line denotes the spin weighted
averaged value of the observed masses of the $1p$ excited $\Lambda_{c}$ states, 
which is 0.33~GeV. 
For the $1d$ states here we call the energetically lower state $1d_{1}$ and the higher $1d_{2}$.
}
\label{fig:EEmodelB}
\end{figure}

Model B has the Gaussian parameter $\beta$ appearing in the size distribution of 
the diquark~\eqref{eq:gauss}.
This parameter is determined so as to reproduce the $1p$ excitation 
energy of the $\Lambda_{c}$ baryon.  In Fig.~\ref{fig:EEmodelB}, we show the
excited energies of the $\Lambda_{c}$ baryon calculated in Model B as functions of 
the Gaussian parameter $\beta$. 
This figure shows that the calculation with $\beta = 1.0$~fm 
reproduces the observed value of the spin average $1p$ excitation energy~\eqref{eq:1pLc}.
Hereafter we take this value of the $\beta$ parameter for Model B. 
The mean squared value of the diquark is calculated as
\begin{equation}
   \langle r^{2} \rangle = \int_{0}^{\infty} \rho^{2} \sigma(\rho; \beta) \rho^{2} d\rho.
\end{equation}
The root mean squared value of the diquark distribution with the Gaussian 
parameter $\beta=1.0$ fm corresponds to 
\begin{equation}
   \sqrt{\langle r^{2}\rangle} = 1.2 \ \textrm{[fm]},
\end{equation}
for $L_{\rho}=0$, which agrees with the Lattice QCD calculation~\cite{Alexandrou}. 
This value implies that the diquark in Model B has an object with a radius of 0.6 fm. 

We fix the absolute value of the effective potential, $V_{0}$, using the mass of the
$\Lambda_{c}$ ground state and obtain
\begin{equation}
   V_{0} = - 0.565\ \textrm{GeV}.
\end{equation}

As explained in Sec.~\ref{sec:model_B}, there are two $1d$ states; 
one is called $1d_{\lambda}$ and has a larger component of the orbital excitation 
of the diquark-quark system, $(L_{\rho}, L_{r}) = (0,2)$, while 
the other is called $1d_{\rho}$ and have a larger component of the rotational 
excitation of the diquark with size, $(L_{\rho}, L_{r}) = (2,0)$.
In Fig.~\ref{fig:EEmodelB}, we call the energetically lower state $1d_{1}$ and the higher $1d_{2}$.
For smaller $\beta$, the lower state is $1d_{\lambda}$, because the rotational excitation costs 
larger energy for a smaller diquark. With larger $\beta$ the rotational excitation energy gets smaller
and level crossing between $1d_{\lambda}$ and $1d_{\rho}$ takes place. Thus, for larger $\beta$ 
the $1d_{\rho}$ state turns to be the lower state. The diquark size $\beta$ at which 
the level crossing takes place depends on the system. 
For the $\Lambda_{c}$ case, the level crossing is found to take place at $\beta = 0.95$ fm. 

\begin{figure}
\centering
\includegraphics[width=0.7\linewidth]{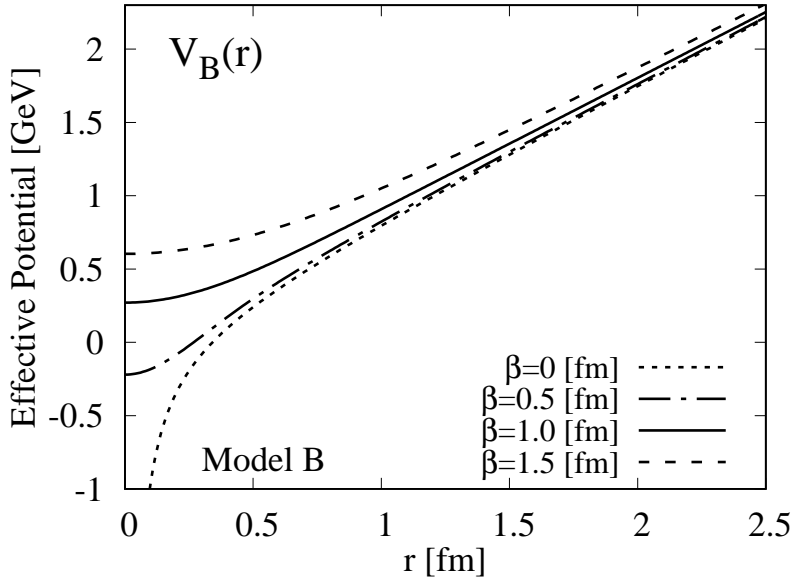}
\caption{ Effective potentials of Model B with $\beta=0, 0.5, 1.0$ and $1.5$ fm. }
\label{fig:VeffB}
\end{figure}

Another characteristic feature of Model B is that the $2s$ and $1d_{\lambda}$ are found
to have a similar energy and tend to degenerate. This is because the effective potential 
between the diquark and heavy quark of Model~B becomes moderate at the origin  
due to the size effect and resembles the harmonic oscillator potential at the short distances
as shown in Fig.~\ref{fig:VeffB}.
The degeneracy of the $2s$ and $1d$ states is one of the general features of the harmonic 
oscillator potential. This tendency is also seen in the quark model calculation of 
the $\Lambda_{c}$ spectrum performed in Ref.~\cite{Yoshida:2015tia}, 
where the $\Lambda_{c}$ baryon is treated as a three-body system of the up, down and 
charm quarks interacting with two-body potentials. The finite distance effect in the $ud$ 
quark system may push up the $2s$ state close to the $1d$ state.

\subsection{Diquark masses}
\label{sec:diquark}

\begin{figure}
\centering
\includegraphics[width=0.7\linewidth]{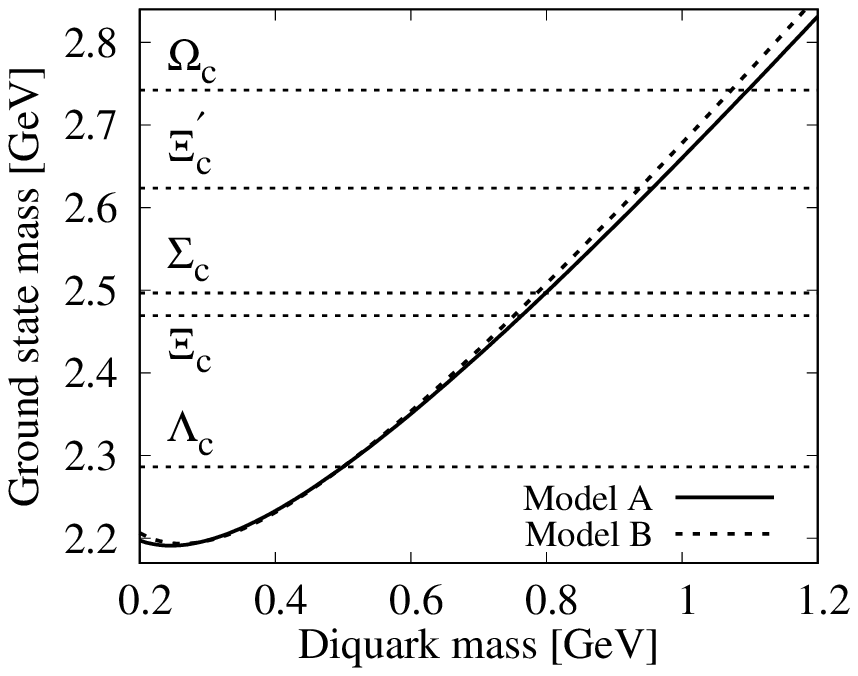}
\caption{Ground state masses of the charmed heavy baryons in Model A and B as 
functions of the diquark mass. The horizontal dotted lines represent the ground state masses of 
the charmed baryons. For the $\Xi^{\prime}_{c}$, $\Sigma_{c}$ and $\Omega_{c}$ baryons, 
we take the spin-weighted average of the masses of the spin 1/2 and 3/2 states. }
\label{fig:1smass}
\end{figure}

In this section, we determine the diquark masses from the ground state masses of the
charmed baryons. The single heavy baryons are characterized by the type of the diquarks
as listed in Table~\ref{tab:diquark}. 
Each diquark has its own spin-flavor content and mass. 
Due to the absence of the interactions inducing the fine structure, 
the spin structure of the charmed baryon does not make difference.
In Fig.~\ref{fig:1smass},
we show the mass of the $1s$ ground state of the singly charmed baryon as 
a function of the diquark mass $m_{d}$ for both Model A and B. 
We show also the charmed baryon masses as the horizontal dotted lines. 
For the $\Sigma_{c}$, $\Xi_{c}^{\prime}$ and $\Omega_{c}$ baryons, which have 
a diquark with spin 1,
we take the spin averaged masses of the spin $1/2^{+}$ and $3/2^{+}$ states 
as their ground states.
From Fig.~\ref{fig:1smass} one can read the value of the diquark mass
for each charmed baryon. The obtained diquark masses are summarized in 
Table~\ref{tab:diquarkmass}. Both models lead to similar values of the diquark masses. 
It is interesting that the mass differences between the $qq$ and $qs$ diquarks 
for spin 0 and spin 1 are 0.25 GeV and 0.15 GeV, respectively. This implies that 
the flavor SU(3) breaking effect is different in the spin 0 and spin 1 diquark.

 \begin{table}[htbp]
  \caption{Diquark masses determined by the ground states of the single charmed baryons
  in units of GeV. The numbers with an asterisk are input values. The diquark is 
  specified by its spin $S$ and flavor $f$. For flavor $f$, $q$ denotes either up or down 
  quark, and $[\cdots]$ and $\{\cdots\}$ 
  imply the antisymmetric and symmetric configurations under the quark exchange,
  respectively. 
  }
  \label{tab:diquarkmass}
   \centering
   \begin{tabular}{c|ccccc} 
   \hline\hline
    $m^{S}_{f}$ & $m^0_{[ud]}$ & $m^{0}_{[qs]}$ & $m^1_{\{qq\}}$ &
    $m^{1}_{\{qs\}}$ & $m^1_{ss}$ \\ 
    \hline 
     Model A & 0.50* & 0.76  & 0.80 & 0.96 & 1.10 \\
     Model B & 0.50* & 0.75  & 0.79 & 0.94 & 1.07 \\
     \hline\hline
    \end{tabular}
\end{table}

\section{Results}
\label{sec:results}

\begin{table}[htb]
\caption{Input model parameters used in this work. The masses $m_{[ud]}^{0}$, $m_{c}$ and $m_{b}$
stand for the masses of the scalar $ud$ diquark, the charm quark and the bottom quark, respectively. 
The coupling constant $\alpha_{s}$ appears in Eqs.~\eqref{eq:potA} and \eqref{eq:intB} 
for the Coulomb part in the quark-diquark interaction of Model A 
and in the quark-quark interaction of Model~B, respectively.
The string tension $k^{\prime}$ is for the quark-quark interaction of Model B,
which reproduces the charmonium and bottonium spectra. 
}
\label{tab:input}
\centering
\begin{tabular}{ccccc}
\hline\hline
 $m_{[ud]}^{0}$ & $m_{c}$ & $m_{b}$ &$\alpha_{s}$ & $k^{\prime}$\\
\hline
  0.5 GeV & 1.5 GeV & 4.0 GeV & 0.4 & 0.9 GeV/fm
\\ \hline\hline
\end{tabular}
\caption{Model parameters of Model A and B. 
The string tension $k$ is for the quark-diquark interaction of Model A~\eqref{eq:potA}, while
the Gaussian parameter $\beta$ determines the diquark size distribution~\eqref{eq:gauss} for Model B.
These parameters are determined by the $\Lambda_{c}$ excitation energy in Sec.~\ref{sec:para}. The absolute value of the effective potential in each model, $V_{0}$,  
is fixed by the mass of the $\Lambda_{c}$ ground state.
}
\label{tab:parameter}
\begin{tabular}{rll}
 \hline\hline
 Model A & $ k = 0.5$ GeV/fm &  $V_{0}= -0.054$ GeV\\
 Model B  & $\beta = 1.0$ fm  &  $V_{0}= -0.565$ GeV
 \\ \hline\hline
\end{tabular}
\end{table}

In this section we show the numerical result of the heavy baryon spectra calculated 
in Model A and B. The parameters introduced as inputs in this work are summarized 
in Table~\ref{tab:input}. The parameters determined with the $\Lambda_{c}$ spectrum 
in the previous section are summarized in Table~\ref{tab:parameter}.
We use the interaction potential with the same parameters for the calculation of the other 
heavy baryon spectra, while the diquark and heavy quark masses are different in each 
system. 

\subsection{$\Lambda_{c}$ excitation spectrum}
\label{sec:Lambdac}
 \begin{figure}[tb]
\centering
    \includegraphics[width=0.7\linewidth]{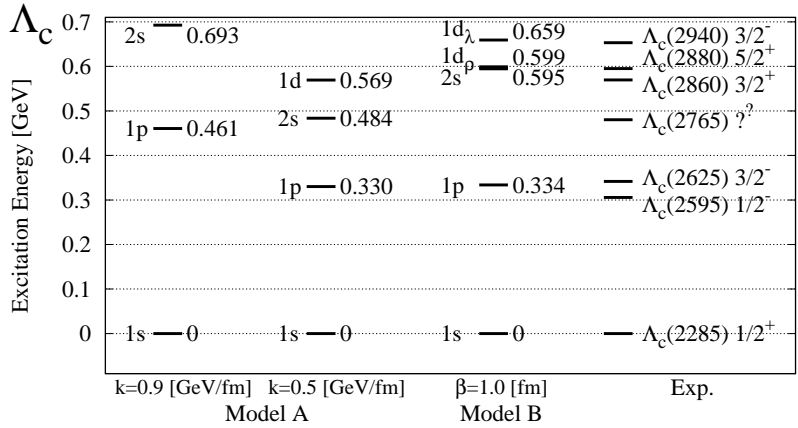}
    \caption{Excitation spectrum of the $\Lambda_{c}$ baryon calculated with Model A and B
    together with the experimental observation. 
    The masses are measured from the ground state in units of GeV. 
    For Model A, we show also the result with $k=0.9$ GeV/fm for comparison. 
    The experimental data are taken from Ref.~\cite{PDG}. 
    }
    \label{result:lambdac}
\end{figure}

First of all, we show the results of the $\Lambda_{c}$ baryon calculated with Models~A
and B in Fig.~\ref{result:lambdac}. 
Both models reproduce the $1p$ excitation energy, as the parameter in each model
is adjusted by the $1p$ excitation energy. 
Here we should compare the calculated values with spin-averaged values of the experimental 
data due to the absence of the spin dependent force in the calculation. 
For Model A, we also show the calculation with
$k=0.9$ GeV/fm that reproduces the charmonium and bottonium spectra well,
which overestimates the $\Lambda_{c}$ excitation energies. 

Figure~\ref{result:lambdac} shows that Model~A with $k=0.5$ GeV/fm reproduces 
the excitation spectrum of $\Lambda_{c}$ well. 
The spin unknown $\Lambda_{c}(2765)$ is reproduced by Model A as a $2s$ state,
while it is not obtained by Model B.  According to Particle Data Group~\cite{PDG} 
the quantum number $\Lambda_{c}(2765)$ state is not known yet including its isospin 
and the observed state could be $\Sigma_{c}(2765)$, while a resent 
experimental analysis~\cite{Belle:2019bab} have found no isospin partners 
and concluded the isospin of $\Lambda_{c}(2765)$ to be zero. 
For further determination of the $\Lambda_{c}(2765)$ quantum number, 
a theoretical analysis~\cite{Arifi:2020ezz} have suggested that the ratio of the decays 
of $\Lambda_{c}(2765)$ to $\Sigma_{c}(2520)\pi$ and $\Sigma_{c}(2455)\pi$ and their angular 
correlations are sensitive to the spin and parity of $\Lambda_{c}(2765)$.
In Model B the $2s$ state appears 100 MeV higher than that of Model A. 
The $2s$ excitation energy is one of the important differences between Models A and B.

Model A obtains the $1d$ state around the energies of the $\Lambda_{c}(2860)$ 
with $J^{P}=(3/2)^{+}$ and $\Lambda_{c}(2880)$ with $J^{P}=(5/2)^{+}$, and thus
these resonances are interpreted as $d$-wave excitations of a quark-diquark system. 
In Model B, the $1d$ states have larger energies than that of Model~A. 
As discussed in Sec.~\ref{sec:ModelBpara}, 
the level crossing between the $1d_{\lambda}$ and $1d_{\rho}$ states takes place at $\beta=0.95$ fm 
for $\Lambda_{c}$ and the $1d_{\rho}$ becomes the lower state for a larger size of the diquark.
Thus, the lower state is found to be $1d_{\rho}$ for $\beta = 1.0$ fm. 
Providing this state close to the observed $\Lambda_{c}(2860)$ and $\Lambda_{c}(2880)$,
Model B interprets the main component of these $\Lambda_{c}$ states 
to be states having a rotational excitation of the diquark with size.

\subsection{Excitation spectra of single heavy baryons}
\label{sec:spectra}

With the diquark masses determined in Sec.~\ref{sec:diquark}, 
we calculate the mass spectra 
of the $\Xi_{c}$, $\Sigma_{c}$, $\Xi_{c}^{\prime}$ and $\Omega_{c}$ charmed baryons
in Model A and B. Once we have determined the potential parameters and the diquark
masses, we have no adjustable parameters in the excitation spectra of these baryons.

\begin{figure}[htbp]
   \centering
    \includegraphics[width=0.7\linewidth]{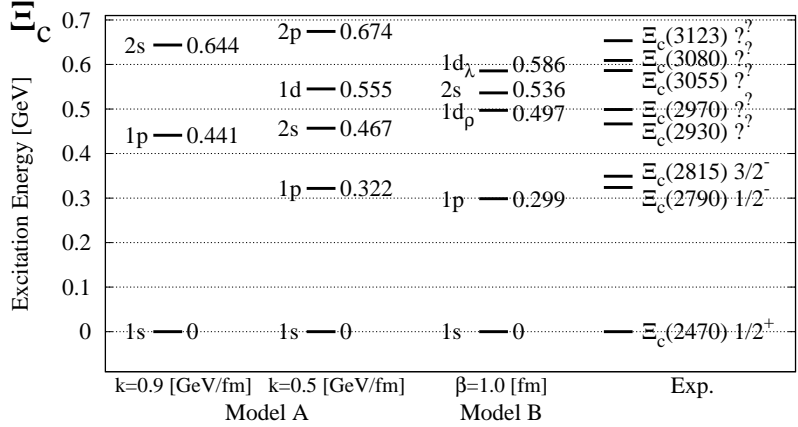}
    \caption{Same as in Fig.~\ref{result:lambdac} but for the $\Xi_{c}$ baryon. 
    Here the $\Xi_{c}^{\prime}$ baryons with $J^{P}=1/2^{+}$ and $3/2^{+}$ are not shown. }
    \label{result:xic}
\end{figure}

First we show the excitation spectra of the $\Xi_{c}$ baryon in Fig.~\ref{result:xic}
together with the experimental observation. In this figure we do not show the
$\Xi_{c}^{\prime}$ baryons with spin 1/2 and 3/2, because we suppose  
these baryons to have a $qs$ diquark with spin 1. 
We find that both models reproduce the $1p$ excitation energy, although Model B 
underestimates the $1p$ energy a bit, which may be reproduced by fine-tuning
the model parameters. Thus, we conclude that both models work for the $\Xi_{c}$ baryon.
Similarly to the $\Lambda_{c}$ spectrum,
there is the characteristic difference between Models A and B in the higher excited
states. In particular, 
the $2s$ state in Model B is higher than that in Model A. The two $1d$ states in Model B
appears close to the $2s$ state. For $\Xi_{c}$ the level crossing between $1d_{\lambda}$ and $1d_{\rho}$ is found to take place at $\beta=0.77$~fm.
There are several excited states observed in experiments, of which spin-parity
is not determined yet. Some of them should be described by the spin 1 $qs$ diquark,
and such states are categorized into $\Xi_{c}^{\prime}$ in this work. 
In order to identify the structure of these baryons,
definitely the information on the spin-parity is necessary. 

\begin{figure}[htbp]
    \centering
    \includegraphics[width=0.7\linewidth]{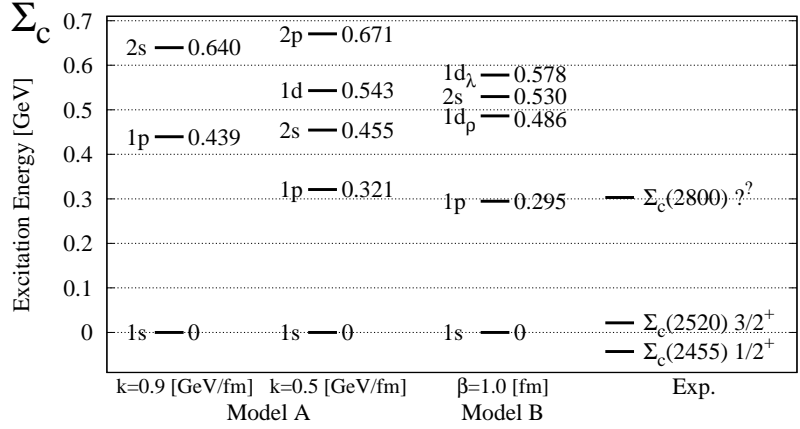}
    \caption{Same as in Fig.~\ref{result:lambdac} but for the $\Sigma_{c}$ baryon. 
    The excitation energy of the $\Sigma_{c}(2800)$ state is measured from 
    the spin average of the masses of $\Sigma_{c}(2455)$ and $\Sigma_{c}(2520)$.  }
    \label{result:sigmac}
\end{figure}

\begin{figure}[htbp]
    \centering
    \includegraphics[width=0.7\linewidth]{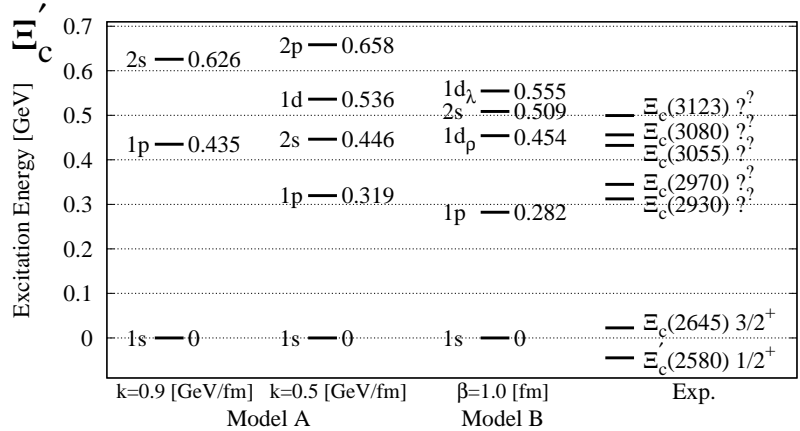}
    \caption{Same as in Fig.~\ref{result:lambdac} but for the $\Xi_{c}^{\prime}$ baryon. 
    The excitation energies of the observed states are measured from the spin average 
    of the masses of $\Xi^{\prime}_{c}(2580)$ and $\Xi^{\prime}_{c}(2645)$. 
    The $\Xi_{c}(2790)$ and $\Xi_{c}(2815)$ baryon with spin $1/2^{-}$ and $3/2^{-}$ 
    are not shown here, because they are assumed as excited states of the $\Xi_{c}$ baryon
    having the $qs$ diquark with spin 0.
    }
    \label{result:xicpra}
\end{figure}

\begin{figure}[htbp]
    \centering
    \includegraphics[width=0.7\linewidth]{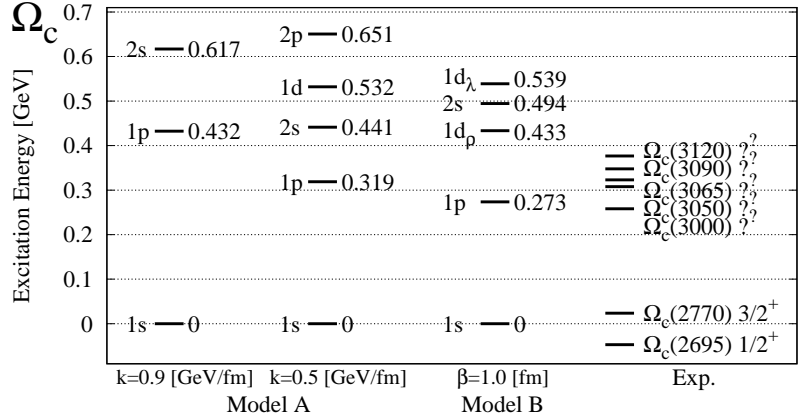}
    \caption{Same as in Fig.~\ref{result:lambdac} but for the $\Omega_{c}$ baryon. 
    The excitation energies of the observed states are measured from the spin average 
    of the masses of $\Omega_{c}(2695)$ and $\Omega_{c}(2770)$. }
    \label{result:omegac}
\end{figure}

In Figs.~\ref{result:sigmac}, \ref{result:xicpra} and \ref{result:omegac},
we show the excitation spectra of the $\Sigma_{c}$, $\Xi_{c}^{\prime}$
and $\Omega_{c}$ baryons, respectively. These baryons are composed 
of a charm quark and a diquark with spin~1. The splitting of the ground states 
is induced by the spin-spin interaction between the charm quark and the spin~1 diquark. 
In the figures the excitation energies of the observed states are measured 
from the spin-weighted average of the ground state masses. 

These figures show that the first excited states of these baryons are 
reproduced well by both models on the whole. 
In Fig.~\ref{result:xicpra}, we do not show the $\Xi_{c}(2790)$ and $\Xi_{c}(2815)$ baryons 
with spin $1/2^{-}$ and $3/2^{-}$, because we assume these resonances to be 
$1p$ excited states of the $\Xi_{c}$ baryon having the $qs$ diquark with spin 0.
The observed $\Xi_{c}$ baryons having the mass of around 2.9 GeV are reproduced as 
a $1p$ state in both models.
For the $p$ state, 
the fine structure splittings are induced by both the spin-spin  
and the spin-orbit interactions. For further comparison we need 
to include such fine structure interactions and to know the 
spin-parity of the observed states. 

It is also interesting to mention that the excited energies obtained in Model A
are less sensitive to the value of the diquark mass than those in Model B. 
In particular, the excitation energies in Model B are reduced as the diquark mass increases
more than those in Model~A. 
In Model B, between two $1d$ states the lower states are found as the $1d_{\rho}$ state
in these baryons. The level crossings of $1d_{\lambda}$ and $1d_{\rho}$ are found 
at $\beta =0.75$~fm for $\Sigma_{c}$, $\beta = 0.68$~fm for $\Xi_{c}^{\prime}$ 
and $\beta=0.63$~fm for $\Omega_{c}$. It is interesting that the level crossing point 
appears at the smaller size parameter for the heavier diquark mass. 
In Table~\ref{tab:charmedbaryon},
we summarize the excitation energies obtained in Models A and B in comparison 
with those of the candidates in the observed baryons. 

\begin{table}[htbp]
  \caption{Excitation energies of the charmed baryons obtained in the calculation of Models~A and B in units of GeV. 
  The observation candidates corresponding to the calculation are also shown.  
  }
  \label{tab:charmedbaryon}
    \centering
    \begin{tabular}{ccD{.}{.}{4}D{.}{.}{4}ccD{.}{.}{4}} 
    \hline\hline
    & $L$ & \multicolumn{1}{c}{Model A}  & \multicolumn{1}{c}{Model B} & candidates & spin 
    & \multicolumn{1}{c}{Exp.}  \\
    \hline
    $\Lambda_{c}$ & $1s$ & 0  & 0   & $\Lambda_{c}$&  $1/2^{+}$ & 0 \\
    & $1p$ & 0.330  & 0.334   & $\Lambda_{c}(2595)$ & $1/2^{-}$ & 0.306 \\
    & & & & $\Lambda_{c}(2625)$ & $3/2^{-}$ & 0.342\\
    & $2s$  & 0.484 &  0.595 & $\Lambda_{c}(2765)$ & $?^{?}$ &0.480 \\ 
    & $1d$  & 0.569 & 0.599\ (\rho) & $\Lambda_{c}(2860)$& $3/2^{+}$ & 0.570\\
    & & & 0.659\ (\lambda) & $\Lambda_{c}(2880)$ & $5/2^{+}$ & 0.595 \\
    \hline
    $\Xi_{c}$ & $1s$ & 0  & 0   & $\Xi_{c}$&  $1/2^{+}$ & 0 \\
    & $1p$ & 0.322  & 0.299   & $\Xi_{c}(2790)$ & $1/2^{-}$ & 0.324 \\
    & & & & $\Xi_{c}(2815)$ & $3/2^{-}$ & 0.349\\
    & $2s$  & 0.467 &  0.536 &  &  & \\ 
    & $1d$  & 0.555 & 0.497\ (\rho)& &  & \\
    & & & 0.586\ (\lambda) & & & \\
    \hline
    $\Sigma_{c}$ & $1s$ & 0  & 0   & $\Sigma_{c}(2455)$&  $1/2^{+}$ & -0.043 \\
    & & & & $\Sigma_{c}(2520)$ & $3/2^{+}$ & 0.022  \\
    & $1p$ & 0.321  & 0.295   & $\Sigma_{c}(2800)$ & $?^{?}$ & 0.303 \\
    & $2s$  & 0.455 &  0.530 &  &  & \\ 
    & $1d$  & 0.543 & 0.486\ (\rho) & &  & \\
    & & & 0.578\ (\lambda) & & & \\
    \hline
    $\Xi_{c}^{\prime}$ & $1s$ & 0  & 0   & $\Xi_{c}^{\prime}$&  $1/2^{+}$ & -0.045 \\
    & & & & $\Xi_{c}(2645)$ & $3/2^{+}$ & 0.022 \\
    & $1p$ & 0.319  & 0.282   & $\Xi_{c}(2930)$ & $?^{?}$ & 0.312 \\
    & & & & $\Xi_{c}(2970)$ & $?^{?}$ & 0.345\\
    & $2s$  & 0.446 &  0.509 &  &  & \\ 
    & $1d$  & 0.536 & 0.454\ (\rho) & &  & \\
    & & & 0.555\ (\lambda) & & & \\
    \hline
    $\Omega_{c}^{\prime}$ & $1s$ & 0  & 0   & $\Omega_{c}$&  $1/2^{+}$ & -0.047 \\
    & & & & $\Omega_{c}(2770)$ & $3/2^{+}$ & 0.024 \\
    & $1p$ & 0.319  & 0.273   & $\Omega_{c}(3000)$ & $?^{?}$ & 0.258 \\
    & & & & $\Omega_{c}(3050)$ & $?^{?}$ & 0.308\\
    & & & & $\Omega_{c}(3065)$ & $?^{?}$ & 0.323\\
    & & & & $\Omega_{c}(3090)$ & $?^{?}$ & 0.348\\
    & & & & $\Omega_{c}(3120)$ & $?^{?}$ & 0.377\\
    & $2s$  & 0.441 &  0.494 &  &  & \\ 
    & $1d$  & 0.532 & 0.433\ (\rho) & &  & \\
    & & & 0.539\ (\lambda) & & & 
    \\ \hline\hline
    \end{tabular}
\end{table}

\begin{figure}[htbp]
    \centering
    \subfigure{
    \includegraphics[width=0.485\linewidth]{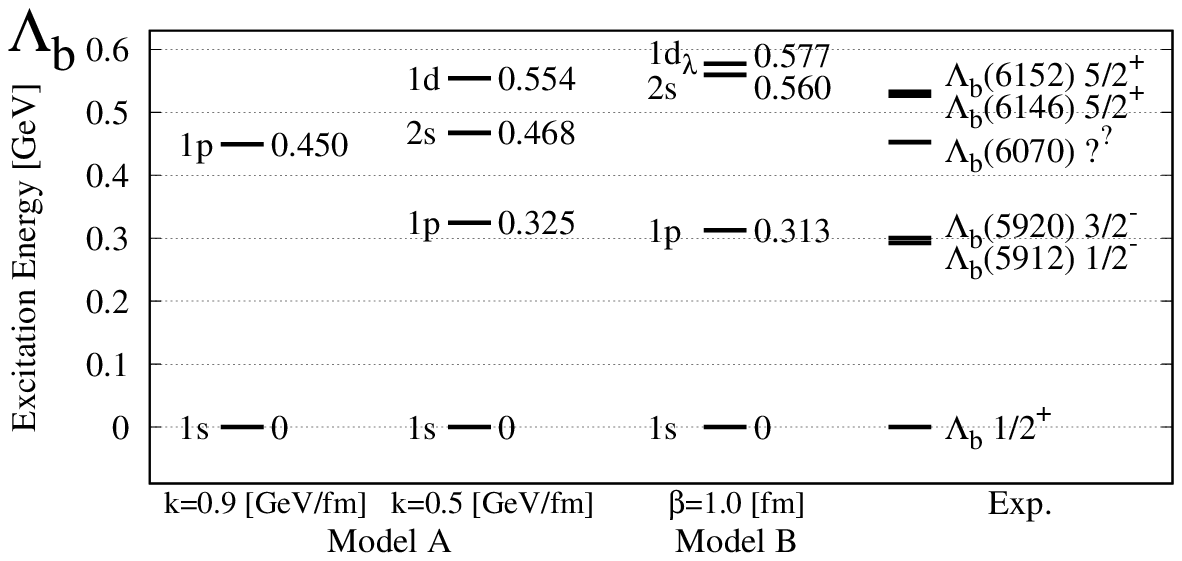}  }
    \subfigure{
    \includegraphics[width=0.485\linewidth]{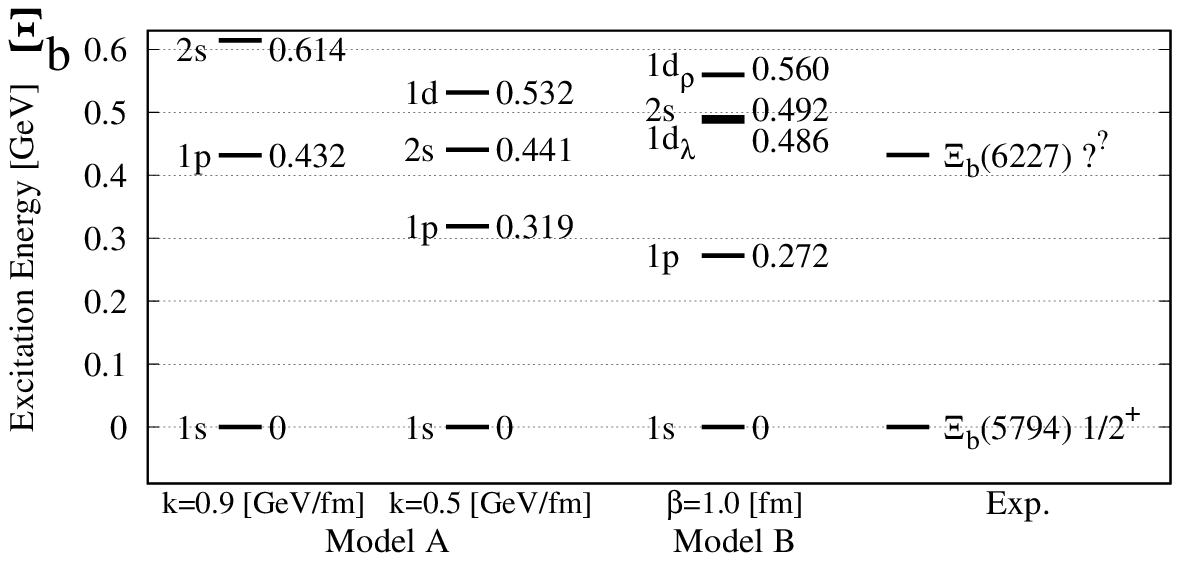}}
    \subfigure{
    \includegraphics[width=0.485\linewidth]{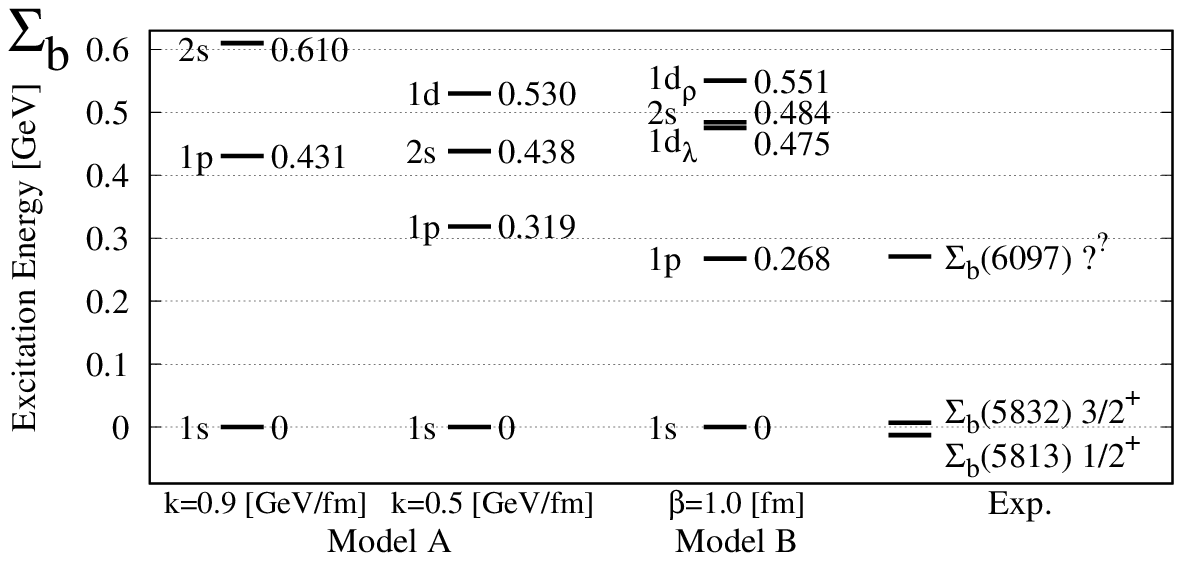}}
    \subfigure{
    \includegraphics[width=0.485\linewidth]{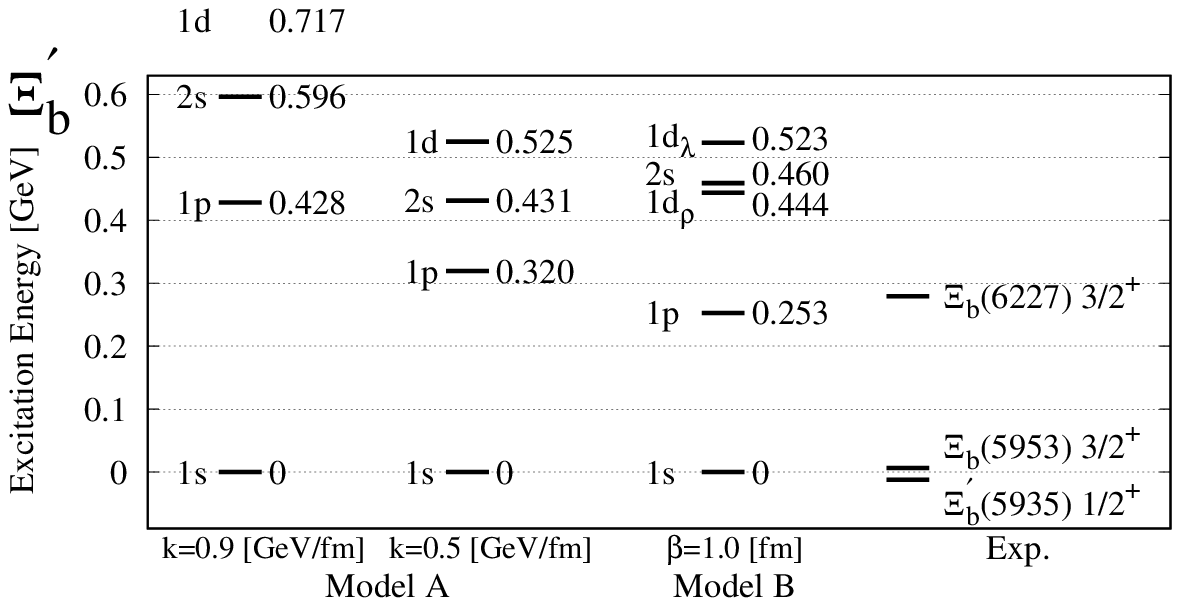}}
    \subfigure{
    \includegraphics[width=0.485\linewidth]{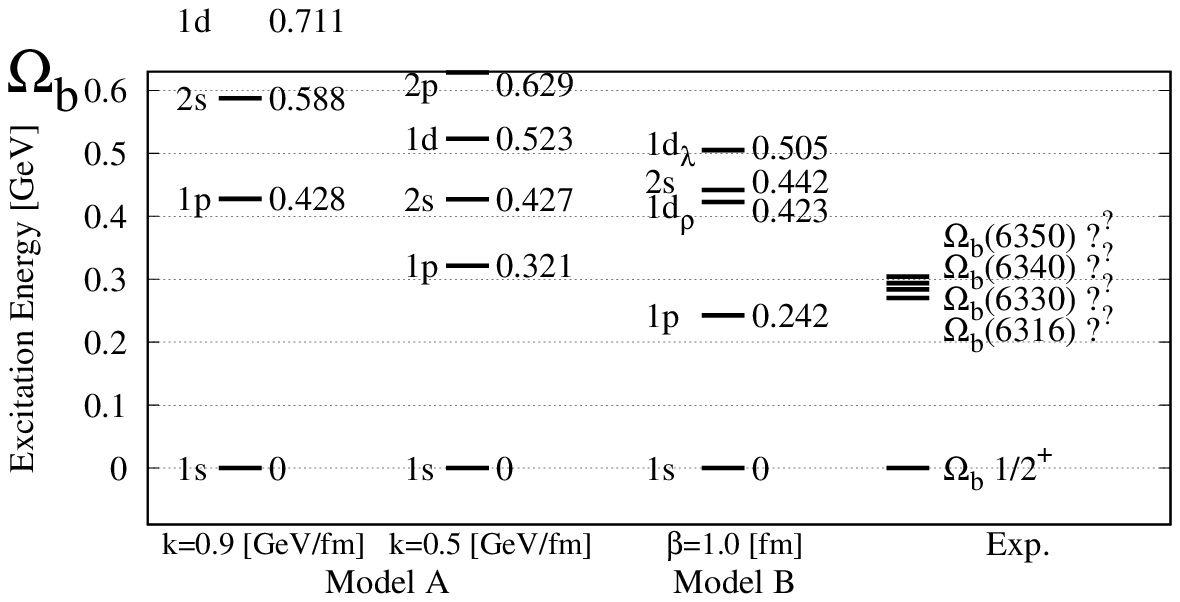}}
    \caption{Excitation spectra of the heavy baryons with one bottom quark calculated
    with Model A and B together with the experimental observation. 
    The masses are measured from the ground states in units of GeV.
    For the ground states of the $\Sigma_{b}$ and $\Xi_{b}^{\prime}$ baryons, 
    the spin-weighted average of the masses of the states with spin $1/2^{+}$
    and $3/2^{+}$ are taken. 
    For Model A, we show also the result with $k=0.9$ GeV/fm for comparison. 
    The experimental data are taken from Ref.~\cite{PDG}. 
    }
    \label{fig:bottom}
\end{figure}

In Fig.~\ref{fig:bottom} and Table~\ref{tab:bottomedbaryon}, 
we show the excitation spectra of the heavy baryons
with one bottom quark in the same way as the charmed baryons. Here 
we assume the bottom quark mass to be 4.0 GeV. All of the model parameters 
are already fixed, and the excitation energies of the bottomed baryons 
are calculated without adjustable parameters. 
The bottom quark mass used here may be smaller than one used for phenomenological
studies, for instance in Ref.~\cite{Karliner:2014gca}. We have also performed a calculation with   
the bottom quark mass 5.0 GeV and found that the results do not change much. 
The differences in the excitation energies are less than 2\%. 
This is because the reduced masses of the bottom quark and the diquark 
are very similar in both cases for such heavy quark masses.

The first excitation energies of the $\Lambda_{b}$ and $\Sigma_{b}$ baryons
are found to be reproduced well in both models. For the heavier bottom baryons,
the excitation energies are reduced generally. Nevertheless, the excitation 
energies in Model A are less sensitive to the diquark mass and the $1p$ excitation 
energies are slightly enhanced in the $\Xi_{b}^{\prime}$ and $\Omega_{b}$ baryons. 
We find again the characteristic difference between two models in the value
of the $2s$ excitation energy. Model~B predicts larger $2s$ excitation energies 
and the $2s$ state almost degenerates with 
one of the $1d$ states.
In Model B, unlike the charmed baryons, the level crossing does not take place 
for $\Lambda_{b}$, $\Xi_{b}$ and $\Sigma_{b}$, while it takes place twice
at $\beta = 0.85$~fm and $\beta = 1.03$~fm for $\Xi_{b}^{\prime}$ and
at $\beta = 0.75$~fm and $\beta = 1.03$~fm for $\Omega_{b}$. 
{For further comparison, we need the detailed information of the 
quantum numbers of the observed states and the observation of missing states. }

\begin{table}[htbp]
  \caption{Excitation energies of the bottomed baryons obtained in the calculation of Model A and B in units of GeV. 
  The observation candidates corresponding to the calculation are also shown.  
  }
  \label{tab:bottomedbaryon}
  \centering
    \begin{tabular}{ccD{.}{.}{4}D{.}{.}{4}ccD{.}{.}{4}} 
    \hline\hline
    & $L$ & \multicolumn{1}{c}{Model A}  & \multicolumn{1}{c}{Model B} & candidates & spin 
    & \multicolumn{1}{c}{Exp.}  \\
    \hline
    $\Lambda_{b}$ & $1s$ & 0  & 0   & $\Lambda_{b}$&  $1/2^{+}$ & 0 \\
    & $1p$ & 0.325  & 0.313   & $\Lambda_{b}(5912)$ & $1/2^{-}$ & 0.293 \\
    & & & & $\Lambda_{b}(5920)$ & $3/2^{-}$ & 0.300\\
    & $2s$  & 0.468 &  0.560 & $\Lambda_{b}(6070)$  &$?^{?}$  &0.453 \\ 
    & $1d$  & 0.554 & 0.577\ (\lambda) & $\Lambda_{b}(6146)$ & $3/2^{+}$  & 0.527 \\
    & & &0.650\ (\rho) & $\Lambda_{b}(6152)$ & $5/2^{+}$ & 0.533  \\
    \hline
    $\Xi_{b}$ & $1s$ & 0  & 0   & $\Xi_{b}$&  $1/2^{+}$ & 0 \\
    & $1p$ & 0.319  & 0.272   &  &  & \\
    & $2s$  & 0.441 &  0.492 &  &  & \\ 
    & $1d$  & 0.532 & 0.486\ (\lambda) & &  & \\
    & & & 0.560\ (\rho) & & & \\
    \hline
    $\Sigma_{b}$ & $1s$ & 0  & 0   & $\Sigma_{b}$&  $1/2^{+}$ & -0.013 \\
    & & & & $\Sigma_{c}^{*}$ & $3/2^{+}$ & 0.006  \\
    & $1p$ & 0.319  & 0.268   & $\Sigma_{b}(6097)$ & $?^{?}$ & 0.271 \\
    & $2s$  & 0.438 &  0.484 &  &  & \\ 
    & $1d$  & 0.530 & 0.475\ (\lambda) & &  & \\
    & & & 0.551\ (\rho) & & & \\
    \hline
    $\Xi_{b}^{\prime}$ & $1s$ & 0  & 0   & $\Xi_{b}^{\prime}(5935)$&  $1/2^{+}$ & -0.013 \\
    & & & & $\Xi_{b}(5945)$ & $3/2^{+}$ & 0.006 \\
    & $1p$ & 0.320  & 0.253   & $\Xi_{b}(6227)$ & $?^{?}$ & 0.279 \\
    & $2s$  & 0.431 &  0.460&  &  & \\ 
    & $1d$  & 0.525 & 0.444\ (\rho) & &  & \\
    & & & 0.523\ (\lambda) & & & \\
    \hline
    $\Omega_{b}^{\prime}$ & $1s$ & 0  & 0   & $\Omega_{b}$&  $1/2^{+}$ & 0 \\
    & $1p$ & 0.321  & 0.242   & $\Omega_{b}(6316)$ & $?^{?}$  & 0.270 \\
    &  &  &   & $\Omega_{b}(6316)$ & $?^{?}$  & 0.270 \\
    &  &  &   & $\Omega_{b}(6330)$ & $?^{?}$  & 0.284 \\
    &  &  &   & $\Omega_{b}(6340)$ & $?^{?}$  & 0.294 \\
    &  &  &   & $\Omega_{b}(6350)$ & $?^{?}$  & 0.304 \\
    & $2s$  & 0.427 &  0.442 &  &  & \\ 
    & $1d$  & 0.523 & 0.423\ (\rho)& &  & \\
    & & & 0.505\ (\lambda)& &  & 
    \\ \hline\hline
    \end{tabular}
\end{table}

\subsection{Regge trajectories}
\label{sec:Regge}

Model A has a smaller string tension than that of our common knowledge, while 
in Model B the string tension that reproduces the heavy quarkonia is used. 
It is interesting to visualize the difference of the global features of the quark-diquark 
interaction. For this purpose, we show the Regge trajectories of the heavy baryons~\cite{Ebert:2011kk,Shah:2016nxi}.
We take the Regge trajectories
of the spin $J$ and mass squared $M^{2}$ given as
\begin{equation}
   J = a M^{2} + a_{0}. \label{eq:regge}
\end{equation}
In our calculation without the fine structure interactions, the total spin $J$ corresponds 
to the relative angular momentum $L$ for the heavy quark and diquark. 
For the absolute values of the bottom baryon masses, we fix the potential parameter $V_{0}$
so as to reproduce the observed ground state $\Lambda_{b}$ mass. Then we obtain 
$V_{0} = 0.824$ GeV for Model A and $V_{0} = 0.310$ GeV for Model B.
We take the lower state of two $1d$ states in the calculation of Model B. 

\begin{figure}[htbp]
    \centering
    \includegraphics[width=0.7\linewidth]{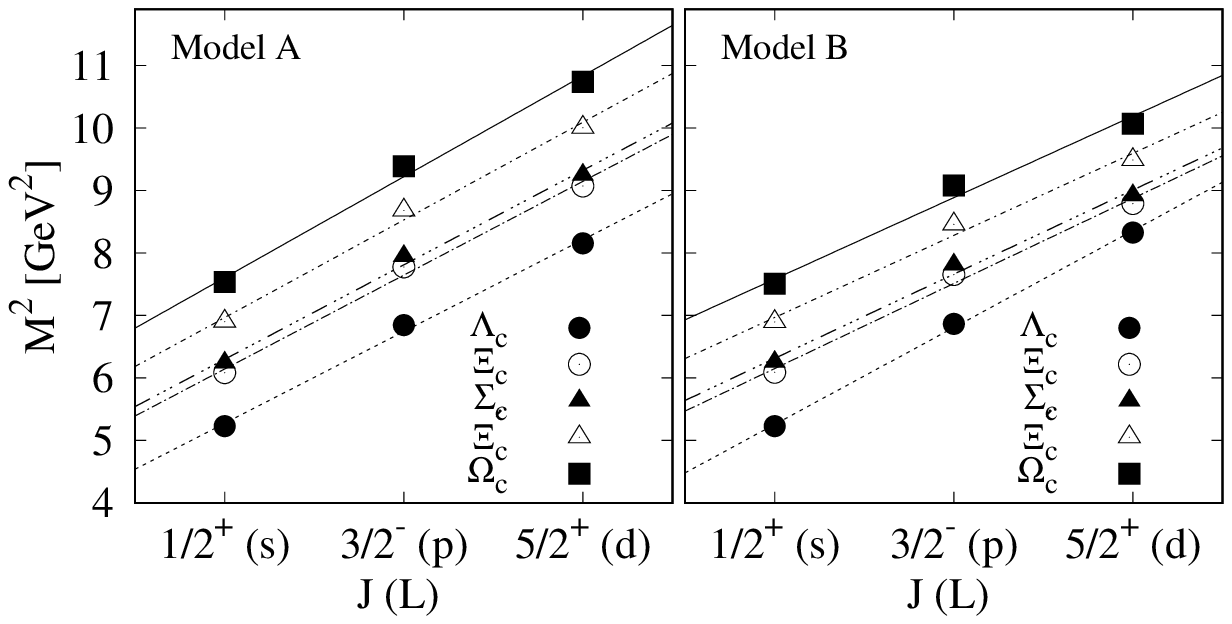}
    \caption{Regge trajectories of charmed baryon masses calculated by
    Model A (left) and Model B (right) as $ J = a M^{2} + a_{0}$. The values of the 
    slop and intersect parameters $a$ and $a_{0}$ are listed Table~\ref{tab:reggeJM}.  }
    \label{fig:reggeJMcharm}
\end{figure}

\begin{figure}[htbp]
    \centering
    \includegraphics[width=0.7\linewidth]{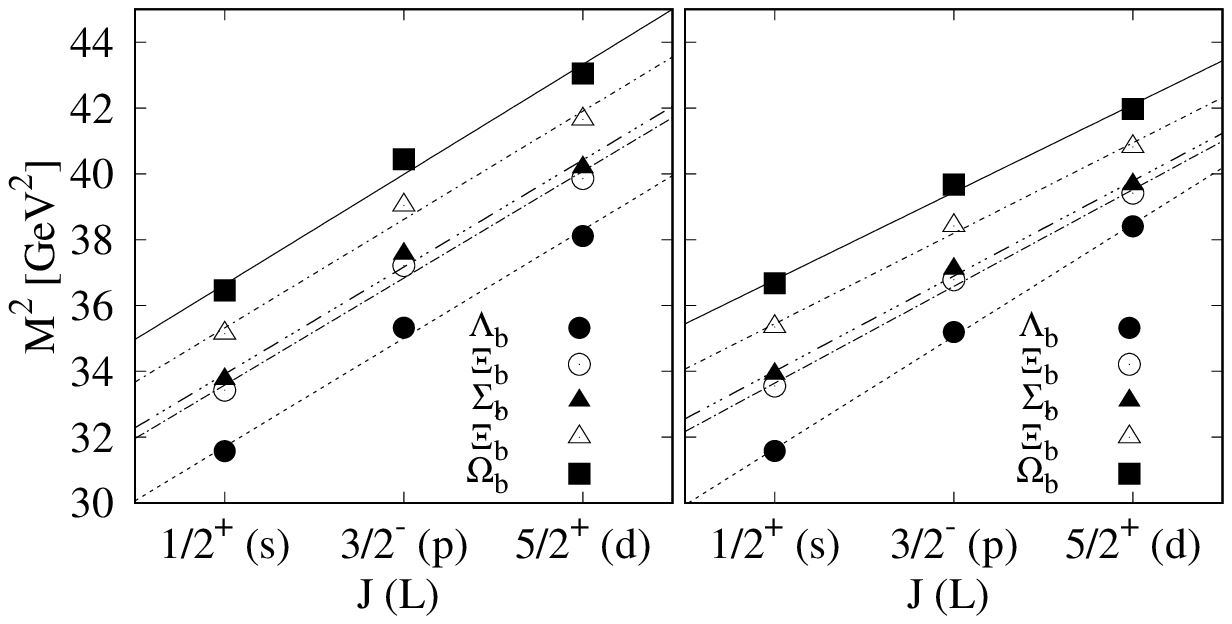}
    \caption{Regge trajectories of bottomed baryon masses calculated by
    Model A (left) and Model B (right) as $ J = a M^{2} + a_{0}$. The values of the 
    slop and intersect parameters $a$ and $a_{0}$ are listed Table~\ref{tab:reggeJM}.  }
    \label{fig:reggeJMbottom}
\end{figure}

We fit the calculated baryon spectra using the Regge trajectory~\eqref{eq:regge}.
The obtained slope and intersect parameters, $a$ and $a_{0}$, are listed 
in Table \ref{tab:reggeJM} for the charmed and bottomed baryons, and 
fitted trajectories are shown in Fig.~\ref{fig:reggeJMcharm} for 
the charmed baryons and Fig.~\ref{fig:reggeJMbottom} for the bottomed baryons. 
It is interesting to find that the slope parameter $a$ obtained in Model~A
for the $\Lambda_{c}$ baryon has a slightly larger value than that of the $\Omega_{c}$
while Model~B has the opposite tendency. The flavor dependence of the 
slope parameter is opposite in Models~A and B.

\begin{table}[htbp]
  \caption{ Fitted values of the slope and intersect parameters given in Eq.~\eqref{eq:regge}
  for the charmed baryons (upper table) and the bottomed baryons (lower table). 
  }
  \label{tab:reggeJM}
  \centering
    \begin{tabular}{ccccc} 
    \hline\hline
     & \multicolumn{2}{c}{Model A}  & \multicolumn{2}{c}{Model B} \\
    & $a$ [GeV$^{2}$] & $a_{0} $    & $a$ [GeV$^{2}$] & $a_{0}$  \\
    \hline
    $\Lambda_{c}$ & $0.681 \pm 0.042$ & $-3.09 \pm 0.28$ & $0.645 \pm 0.021$ & $-2.89 \pm 0.02$ \\
    $\Xi_{c}$ & $0.666 \pm 0.051$ & $-3.59 \pm 0.39$ & $0.734 \pm 0.067$ & $-4.01 \pm 0.51$ \\
    $\Sigma_{c}$ & $0.661 \pm 0.052$ & $-3.66 \pm 0.41$ & $0.742 \pm 0.073$ & $-4.19 \pm 0.56$ \\
    $\Xi_{c}^{\prime}$ & $0.640 \pm 0.054$ & $-3.96 \pm 0.47$ & $0.761 \pm 0.090$ & $-4.80 \pm 0.75$ \\
    $\Omega_{c}$ & $0.619 \pm 0.056$ & $-4.21 \pm 0.52$ & $0.767 \pm 0.101$ & $-5.31 \pm 0.91$ \\
    \hline
    $\Lambda_{b}$ & $0.304 \pm 0.026$ & $-9.13 \pm 0.92$ & $0.293 \pm 0.010$ & $-8.77 \pm 0.35$ \\
    $\Xi_{b}$ & $0.308 \pm 0.032$ & $-9.83 \pm 1.17$ & $0.340 \pm 0.020$ & $-10.93 \pm 0.73$ \\
    $\Sigma_{b}$ & $0.307 \pm 0.032$ & $-9.92 \pm 1.21$ & $0.343 \pm 0.021$ & $-11.26 \pm 0.79$ \\
    $\Xi_{b}^{\prime}$ & $0.304 \pm 0.035$ & $-10.23 \pm 1.34$ & $0.364 \pm 0.026$ & $-12.39 \pm 0.98$ \\
    $\Omega_{b}$ & $0.300 \pm 0.036$ & $-10.46 \pm 1.45$ & $0.375 \pm 0.028$ & $-13.30 \pm 1.12$ 
    \\ \hline\hline
    \end{tabular}
\end{table}

\subsection{Doubly charmed baryon}
\label{sec:double}

We have obtained a smaller string tension for 
{the systems of a point-like diquark and a heavy quark than those of a quark and an antiquark}
in order to reproduce the excitation energies 
of the single heavy baryons. 
It is interesting to examine whether this conclusion is valid only for the light diquarks 
or a heavy diquark system also has similarly such a smaller string tension.
For this purpose, we calculate the excitation spectra
of the doubly charmed baryon~$\Xi_{cc}$ using the quark-diquark model (Model A)
with the string tensions $k=0.9$~GeV/fm and $k=0.5$~GeV/fm, and we compare 
the excitation energy spectra obtained by these calculations.

\begin{figure}[htbp]
   \centering
    \includegraphics[width=0.7\linewidth]{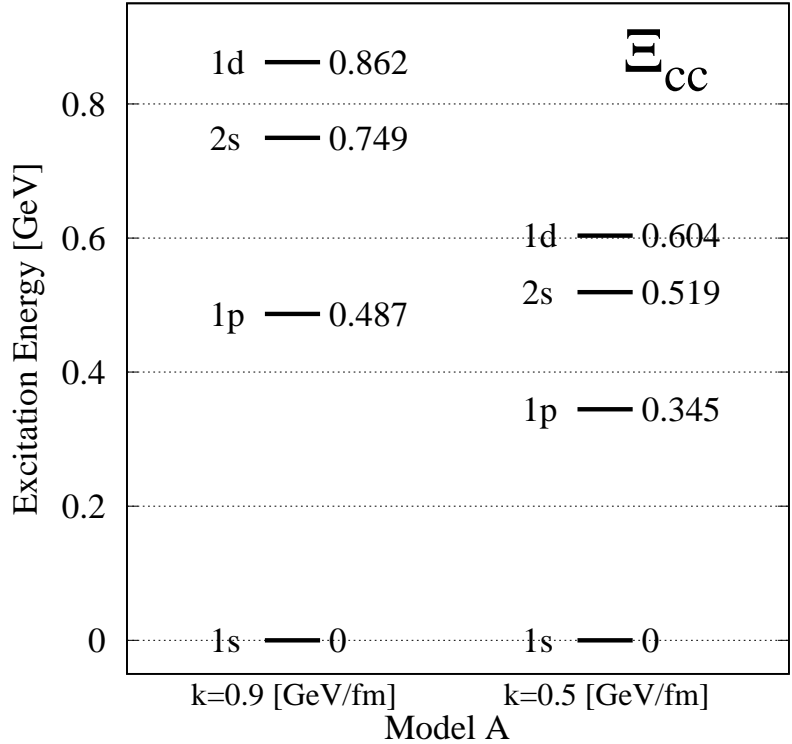}
    \caption{Excitation spectrum of the doubly charmed baryon $\Xi_{cc}$
    calculated by the quark-diquark model. The left and right spectra are 
    obtained using the string tensions $k=0.9$ GeV/fm and $k=0.5$ GeV/fm,
    respectively.}
    \label{fig:Omegacc}
\end{figure}

The $\Xi_{cc}$ is composed of a light quark (up or down quark) and two 
charm quarks. 
In the calculation we treat the two charmed quarks as a point-like diquark.
This might be a good approximation, because the light quark spreads over
more widely than the charm quarks. 
The potential for the quark-diquark system is given by Eq.~\eqref{eq:potA}.
For the string tension we use $k=0.9$~GeV/fm and $k=0.5$~GeV/fm
and $\alpha_{s}=0.4$ for the Coulomb part. The masses of the 
light quark and the charm diquark are assumed to be 
$m_{q} = 0.3$ GeV and $m_{cc} =3.0$~GeV.
The excitation spectra obtained by these calculations are shown in Fig.~\ref{fig:Omegacc}. 
As seen in the figure, the obtained spectra are obiously different:
the excitation energy obtained by using $k=0.9$~GeV/fm 
is about 40\% larger than that obtained with $k=0.5$~GeV/fm. 
Recently the doubly charmed baryon has been found at LHCb~\cite{LHCb:2017iph}, but 
its excited states have not been observed yet. 
It would be very interesting if one could observe 
the first excited state of the $\Xi_{cc}$ baryon and measure its excitation energy.
Such observation would reveal the nature of the two quark correlation.

\section{Summary}
\label{sec:summary}

We have investigated the single heavy quark baryons using 
the quark-diquark models, in which two light quarks are treated as a diquark
and the system is reduced to a two-body problem,
in order to investigate the properties of the diquark constituent in the hadron structure.
We consider two models: In Model~A the diquark is a point-like object 
and the confinement potential for the quark-diquark system
has a smaller string tension $k=0.5$~GeV/fm to reproduce the excitation energies
of the $\Lambda_{c}$ baryon
as suggested in Ref.~\cite{Jido:2016yuv,Kumakawa:2017ffl}. 
In Model~B the two quarks in the diquark have a distance given by the Gaussian 
distribution and the interquark potential is consistent with the
quark-antiquark system in which the string tension is found to be  $k=0.9$ GeV/fm.
The potential for the quark-diquark system is obtained by folding the 
interquark potential obtained with a fixed distance over the Gaussian distribution 
of the distance as given in Eq.~\eqref{eq:potB}.
The size of the Gaussian distribution is determined 
so as to reproduce 
the spin averaged excitation energy of the first excited states of $\Lambda_{c}$
and is found to be $\beta =1.0$~fm which corresponds to a mean root square distance 
$\sqrt{\langle r^{2}\rangle} = 1.2$~fm. 
The scalar $ud$ diquark mass is fixed as 0.5 GeV as an input, while 
the other diquark masses are determined by the ground state masses of
the charmed baryons. 

Model A reproduces well
the spectra of the single heavy baryons on the whole, while
Model~B overestimates the $2s$ state and obtains its energy close 
to the $1d$ states.
This tendency is also found in a three-body calculation of heavy baryons 
in Ref.~\cite{Yoshida:2015tia}.
If Model A is the case, the string tension between the heavy quark and the 
light diquark is as weak as half of that for the quark-antiquark potential 
appearing in mesonic systems, although the color configurations are 
same. 
It would be very interesting if the nature of the $2s$ states of the 
single heavy baryons could be clarified in experimental observations 
because it reveals the properties of the diquark in the heavy baryons. 
We have also examined the Regge trajectories to understand the
global feature of the heavy baryon spectra. We have found  
the difference of two models in the flavor dependence of the slopes
of the trajectories. 
In order to confirm the strength of the quark-diquark interaction, 
we propose to observe the excitation energy of the double charm 
quark $\Xi_{cc}$. 

In conclusion, 
the diquark models reproduce the orbital excitation energies of the single heavy baryons
and the diquarks can work very well as their constituent. 
The investigation of the decay properties of the heavy 
baryons based on this picture will be one of the further confirmations. 
It would be also interesting if one could 
understand the favorable outcomes of the diquark model from much more sophisticated calculations.

\section*{Acknowledgments}
The work of D.J.\ was supported by the Grant-in-Aid for Scientific Research (Nos.\ JP17K05449 and 21K03530) from JSPS.

\let\doi\relax

\end{document}